\documentstyle[twocolumn,eqsecnum,aps,pre]{revtex}

\input epsf.sty

\begin{document}
%\draft
\wideabs{
\title{Surface width scaling in noise reduced
Eden clusters}
\author{M.~T. Batchelor}
\address{Department of Mathematics, School of Mathematical Sciences,\\
Australian National University, Canberra ACT 0200, Australia}
\author{B.~I. Henry and S.~D. Watt}
\address{Department of Applied Mathematics, University of New South
Wales, Sydney NSW 2052, Australia}
\date{\today}
\maketitle

\begin{abstract}
The surface width scaling of Eden A clusters grown
from a single aggregate site on the square lattice
is investigated as a function of the noise reduction parameter.
A two-exponent scaling ansatz is introduced and 
used to fit the results from simulations
covering the range from fully stochastic to the zero-noise limit.

\end{abstract}
}
\narrowtext

\section{Introduction}
The Eden model \cite{E}, which was originally introduced
to model the growth of cell colonies,
is one of the simplest and most widely studied \cite{E,WB,PSHL,M,MW,PR,RP,FV,JB,PR2,FSS,KPZ,MJB,HW,ZS1,ZS2,W,WK,KW,M2,DS,BH}
growth models and has become a paradigm for studying
 self affine fractal geometry and scaling
in the growth of rough surfaces \cite{M3,BS,M4}.
In the simplest variant of the model (Eden A) 
aggregate particles are added one at a time to
randomly selected sites on the surface of a growing cluster
\cite{PSHL,M}.
The aggregate sites and surface sites
are situated at the vertices
of a regular lattice in on-lattice simulations.
The initial cluster is usually taken to be a single aggregate
site in the plane (radial growth) or a line of aggregate sites
in the half plane (substrate
growth).
Most studies of the Eden model have been concerned with
describing the asymptotic properties of the surface of the cluster.
One of the schemes that has been introduced to better reveal
these properties is noise reduction \cite{M3,BS,M4}.
Noise reduction is implemented by
associating a counter (initially set to zero) with each 
of the surface sites
and incrementing the counter by one each time the associated
surface site is selected for growth \cite{WK,KW,M2,DS,BH}.
An aggregate particle is then added at a surface site
when its associated counter reaches a prescribed value $m$.
Increasing values of $m$ lead
to increasingly smoother interfaces.
It is widely believed that
noise reduction reveals asymptotic surface properties
at smaller system sizes without affecting the surface scaling exponents.
This tenant is investigated in this paper for the Eden A model
with growth from a seed on a square lattice.

\section{Surface scaling ansatz}
Extensive studies of Eden growth from a substrate
have identified scaling exponents
$\alpha$ and $z$
relating the surface thickness 
$w$ to the substrate width $L$
and the mean surface height $\langle h \rangle$.
For a cluster with $N$ aggregate particles and
$\cal N$ surface sites this relationship has the form \cite{FV}
\begin{equation}
w\sim L^\alpha f\left(\frac{\langle h \rangle}{L^z}\right)\label{FVscale}
\end{equation}
where the width is defined by
\begin{equation}
w^2=\langle h^2 \rangle- \langle h \rangle^2=
\frac{1}{\cal N}\sum_{i=1}^{\cal N} h_i^2-
\left(\frac{1}{\cal N}\sum_{i=1}^{\cal N} h_i\right)^2\label{Subw}
\end{equation}
and the scaling function $f(x)$ has the properties
\begin{equation} f(x)\propto
\left\{\begin{array}{rl}
        x^{\alpha/z} & \mbox{  for  $x\ll 1$  }\\
\mbox{const}\label{scale} & \mbox{  for $x\gg 1$. } \end{array} \right.
\end{equation}
It follows from the scaling laws,
Eqs. (\ref{FVscale}),(\ref{scale}),
 that for $L$ large the surface of Eden clusters on a 
two-dimensional substrate
is a self-affine fractal with Hurst exponent $\alpha$ and fractal dimension
$2-\alpha$.

Collective evidence from the numerical simulations
and algebraic calculations \cite{KPZ} suggest the
values $\alpha\simeq 1/2$ and $z\simeq 3/2$ (or $\beta\equiv\alpha/z\simeq 1/3$)
for two-dimensional Eden growth on a substrate \cite{M3,BS,M4}.
Although it should be pointed out that the numerical results are not
definitive due to finite-size effects and the algebraic results
may not be entirely applicable since they are based
on a continuum model that includes surface relaxations.

For an Eden cluster growing in a circular geometry
 with $N$ aggregate particles, $\cal N$ surface sites
and an average radius $\langle R \rangle$ (which grows linearly with time)
 the interface width
\begin{equation}
w^2=\langle R^2 \rangle-\langle R \rangle^2
=\frac{1}{\cal N}\sum_{i=1}^{\cal N} R_i^2-
\left(\frac{1}{\cal N}\sum_{i=1}^{\cal N} R_i\right)^2\label{radw}
\end{equation}
is expected to scale as 
\begin{equation}
w\sim {\langle R \rangle}^\beta \label{radscale}.
\end{equation}
Numerical simulations of Eden A clusters
on the square lattice with up to $N\approx 10^5$
aggregate particles again suggest $\beta\simeq 1/3$.
However, very large simulations for
$N\approx 10^7$, \cite{FSS} and $N\approx 10^9$
\cite{ZS1} reveal the increasingly dominant effects of lattice anisotropy
where eventually it is expected \cite{ZS1} that $w\sim \langle R \rangle$.
In this paper we have carried out brute force
simulations of the Eden A model on the square lattice
starting from a single aggregate site over a range of $m$ from
the fully stochastic limit $m=1$ to the zero-noise limit $m\to\infty$
\cite{BH}.
The brute force calculations 
avoid the possibility of numerical bias from
e.g., quadrant boundary effects \cite{FSS} or multiply selected surface sites
\cite{ZS1}.
The results of the simulations are shown to be consistent
with a two-exponent scaling ansatz of the form
\begin{equation} w(N,m)\sim
\left\{\begin{array}{rl}
        a(m) N^{\frac{1}{6}} & \mbox{  for  $N\ll N^*(m)$  }\\
b(m) N^{\frac{1}{2}}\label{ansatz} & \mbox{  for $N\gg N^*(m)$ } \end{array} \right.
\end{equation}
where $N^*(m)$ denotes an empirical cross-over number of aggregate particles
for a given $m$.
This is in agreement with the expectation that noise reduction
does not affect the values of the scaling exponents; however the
cross-over value
\begin{equation}
N^*=\left(\frac{a(m)}{b(m)}\right)^3
\end{equation}
is $m$ dependent.
The two-exponent scaling ansatz can also be written in the
functional form
\begin{equation}
w(N,m)\sim b(m) N^{\frac{1}{2}}g\left(\frac{N}{N^*(m)}\right)\label{ansatz2}
\end{equation}
where
\begin{equation} g(x)=
\left\{\begin{array}{rl}
        x^{-1/3} & \mbox{  for  $x\ll 1$  }\\
1\label{ascale} & \mbox{  for $x\gg 1$. } \end{array} \right.
\end{equation}

\section{Zero-Noise Limit}

In the zero-noise limit $m\to\infty$, Eden A clusters on a square lattice
grow in layers as a compact diamond with
\cite{BH}
\begin{equation}
N_k=2k^2-2k+1,\qquad k=1,2,\ldots \label{Dnos}
\end{equation}
aggregate particles. In this limit there is no stochastic growth
so that
\begin{equation}
\lim_{m\to\infty} N^*(m)\rightarrow 0
\end{equation}
and 
\begin{equation}
\lim_{m\to\infty} w(N,m)\sim b(\infty)N^{\frac{1}{2}}.
\end{equation}
A simple approximation to $b(\infty)$ can be found from the continuum
expression for a diamond in polar co-ordinates:
\begin{equation}
R(\theta)=\frac{\sqrt{N}}{\sqrt{2}\left(|\sin \theta |+|\cos \theta |\right)}.
\end{equation}
The averages over $\theta$ can be calculated exactly
yielding
\begin{equation}
w\sim 
\sqrt{\frac{1}{\pi}-\frac{4\left(\tanh^{-1} 
\frac{1}{\sqrt{2}} \right)^2}{\pi^2}}N^{\frac{1}{2}}
 \label{Dscale}
\end{equation}
and hence $b(\infty)\approx 0.05896\ldots$.

\section{Numerical Results}

The results described in this section summarize
data from our numerical simulations
of ensembles of Eden A clusters starting from a single seed
on the square lattice.
Each ensemble
consists of one hundred Monte-Carlo simulations
of the Eden A  model
for a fixed value of the noise-reduction parameter $m$.
The surface width, Eq. (\ref{radw}), is averaged over the ensemble copies
to obtain the surface width as a function of $N$ for a given $m$.

Figure 1 shows plots of the ensemble averaged surface width versus
the number of aggregate particles
(using a log-log scale) for
 $m$ at one unit intervals in the range
$m\in[1,64]$ and for $m\to\infty$ (dashed line).
The curves for increasing values of $m$ are
from right to left on the right hand side of the plot and
upper to lower on the left hand side of the plot.
The peaks in the sawtooth pattern for large $m$ and small $N$
occur at exact `diamond numbers', Eq. (\ref{Dnos}).
The surface width data was fit to the two-exponent scaling anstaz, 
Eq. (\ref{ansatz}).
In figure 2 the best fit estimates 
for a) $a(m)$ and b) $b(m)$ are plotted against
 $m$ at one unit intervals in the range
$m\in[1,64]$. 
 Figure 2 b) also shows (dashed line) the
asymptotic value of $b(\infty)$ obtained from the 
calculation in the zero-noise limit, Eq. (\ref{Dscale}).
The surface profile of very large Eden clusters at $m=1$ is slightly anisotropic
and well fit (in the first quadrant) by
\begin{equation}
R(\theta)=\langle R \rangle+A\cos 4\theta \label{aniso}.
\end{equation}
where $\langle R\rangle\approx \sqrt{\frac{N}{\pi}}$ is the average radius 
of the cluster and $A$ is the amplitude
of the anisotropy (about one percent of $\langle R \rangle$ \cite{ZS1}).
The anisotropic profile, Eq. (\ref{aniso}), has a surface width 
\begin{equation}
w\sim\frac{A}{\sqrt{2\pi}}N^{\frac{1}{2}}.
\end{equation}
Our value $b(1)\approx .005$ is thus consistent with a slight anisotropy of the
order of $A\approx \pm 1\% \langle R \rangle$.

The functional form of the two-exponent scaling ansatz, Eq. (\ref{ansatz2}),
is clearly revealed in Figure 3 where we have collapsed
the surface width data for all $N$ and $m$ values onto
a single curve by plotting
$$
\frac{w(N,m)}{b(m)N^{1/2}} \quad\mbox{versus}\quad \frac{N}{N^*(m)}.
$$

\section{Discussion}

In this Brief Report we showed that the surface width of 
noise-reduced Eden A clusters
grown from a seed on a square lattice scales with the number of 
aggregate particles according to a 
a two-exponent relation.
The exponents $1/6$ for $N<N^*$ and $1/2$ for
$N>N^*$ were found to be independent of the noise
reduction parameter $m$ but the crossover value $N^*$
was found to decrease monotonically with $m$.
These results support the tenants that: i) noise reduction
does not affect the scaling exponents in Eden-like growth models
and ii) increasing noise reduction decreases the size of clusters
needed for observing the limiting large $N$ scaling behaviour.
In particular, provided the scaling coefficients
do not vanish in the limit $m\rightarrow\infty$, the large $N$
 scaling exponents could be found
rather simply from exact (algebraic or numerical) calculations
in the zero-noise limit.
On the other hand intermediate
scaling results from finite $m$
simulations would have to be interpreted with some
caution particularly in cases where the growth is characterized
by  multi-exponent scaling laws and
multiple $m$ dependent cross-over values.
It is anticipated that a similar two-exponent scaling relation
to Eq. (\ref{ansatz2}) with the same two scaling exponents but 
different coefficients $a(m)$ and $b(m)$
could be used to characterize on-lattice
(e.g., square, triangular or honeycomb) growth
of other variants of the Eden model
(e.g., Eden B and Eden C).

\acknowledgments
This work was supported by the Australian Research Council.

\begin{figure}
\vskip -2cm
\centerline{
\epsfxsize=3.5in
\epsfbox{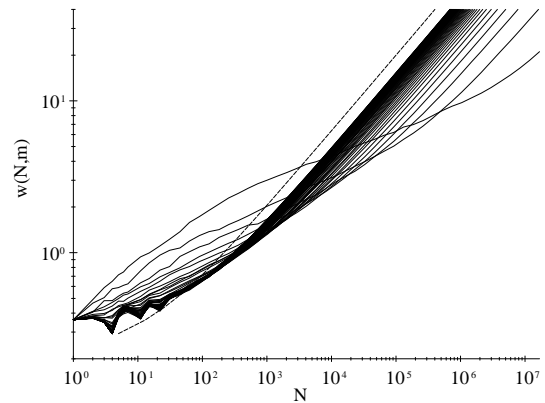}}
\vskip -2cm
\caption{Plots of the surface width versus the number of aggregate
particles for different values of the noise reduction parameter
at one unit intervals in the range $m\in[1,64]$
and for the zero-noise limit (dashed curve).
Note the logarithmic scale on each of the axes.}
\label{fig1}
\end{figure}

\begin{figure}
\centerline{
\epsfxsize=3.5in
\epsfbox{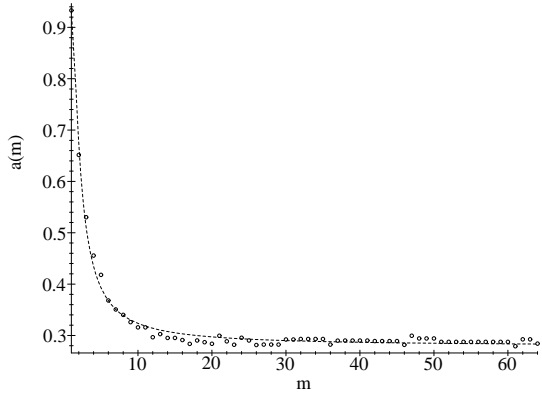}}
\vskip -5cm
\centerline{
\epsfxsize=3.5in
\epsfbox{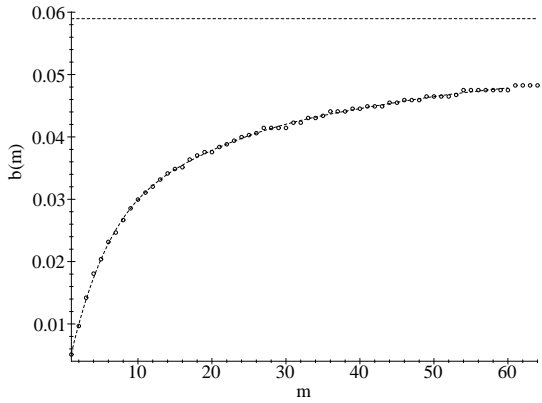}}
\vskip -2cm
\caption{Plots of the surface width scaling coefficients in
Eq. (\ref{ansatz}) as a function
of the noise reduction parameter:
a) $a(m)$ versus $m$ and b) $b(m)$ versus $m$.}
\label{fig2}
\end{figure}

\begin{figure}
\centerline{
\epsfxsize=4.0in
\epsfbox{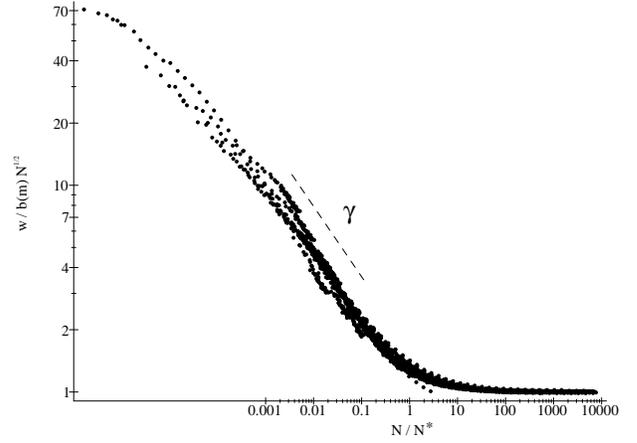}}
\vskip -2cm
\caption{Plots of
$\frac{w(N,m)}{b(m)N^{1/2}}$ versus $\frac{N}{N^*(m)}$
for each integer value of $m$ in the range $[1,64]$.
The surface width $w(N,m)$ is averaged over one hundred different Monte-Carlo
simulations at that value of $N$ and $m$.
The dashed line has a slope $\gamma=-1/3$.
Note the logarithmic scale on each of the axes.}
\label{fig3}
\end{figure}

\end{document}